\documentclass[aps,prl,twocolumn,showpacs,superscriptaddress,groupedaddress,nofootinbib,nobibnotes]{revtex4-1}

\usepackage[english]{babel}
\usepackage[latin1]{inputenc}

\usepackage{longtable}

\usepackage{graphicx} 
\usepackage{amssymb}

\usepackage{multirow}  
\usepackage{tabularx}

\usepackage[bookmarks=true, colorlinks=true,urlcolor=black, citecolor=black, linkcolor=black, bookmarksnumbered=true, hypertexnames=false, breaklinks=true]{hyperref} 
\usepackage[hyperref]{xcolor}
		\definecolor{light-blue}{rgb}{0.1,0.1,0.6}

\usepackage[nolist,nohyperlinks]{acronym}

\usepackage{bm}

\usepackage{color}

\begin{document}

\title{Study of e$^+$e$^-$ $\boldmath \rightarrow \text{p}  \overline{\text{p}}\pi^0$ in the Vicinity of the $\boldmath \psi\text{(3770)}$~}
\date{\today}
\author{
M.~Ablikim$^{1}$, M.~N.~Achasov$^{8,a}$, X.~C.~Ai$^{1}$, O.~Albayrak$^{4}$, M.~Albrecht$^{3}$, D.~J.~Ambrose$^{41}$, F.~F.~An$^{1}$, Q.~An$^{42}$, J.~Z.~Bai$^{1}$, R.~Baldini Ferroli$^{19A}$
% ...
\\
\vspace{0.2cm}
(BESIII Collaboration)\\
\vspace{0.2cm} {\it
$^{1}$ Institute of High Energy Physics, Beijing 100049, People's Republic of China\\
% ...
$^{3}$ Bochum Ruhr-University, D-44780 Bochum, Germany\\
$^{4}$ Carnegie Mellon University, Pittsburgh, Pennsylvania 15213, USA\\
% ...
$^{8}$ G.I. Budker Institute of Nuclear Physics SB RAS (BINP), Novosibirsk 630090, Russia\\
% ...
$^{19}$ (A)INFN Laboratori Nazionali di Frascati, I-00044, Frascati, Italy; (B)INFN and University of Perugia, I-06100, Perugia, Italy\\
% ...
$^{41}$ University of Rochester, Rochester, New York 14627, USA\\
$^{42}$ University of Science and Technology of China, Hefei 230026, People's Republic of China\\
% ...
$^{a}$ Also at the Novosibirsk State University, Novosibirsk, 630090, Russia\\
% ...
\ \\
}
}

\begin{abstract}
The process $e^+e^-\rightarrow p\overline{p}\pi^0$ has been studied by analyzing
data collected at $\sqrt{s}=3.773$~GeV, at $\sqrt{s}=3.650$~GeV, and during a $\psi(3770)$ line shape scan with the \acs{BESIII} detector at the
\acs{BEPCII} collider. 
The Born cross section of $p \overline{p} \pi^0$ in the vicinity of the $\psi(3770)$ is measured and the Born cross section of $\psi(3770)\rightarrow p \overline{p} \pi^0$ is extracted
considering interference between resonant and continuum production amplitudes. Two solutions with the same probability and a significance of 1.5$\sigma$ are found. The solutions
for the  Born cross section of $\psi(3770)\rightarrow p \overline{p} \pi^0$ are $33.8\pm1.8\pm2.1$~pb and $0.06^{+0.10+0.01}_{-0.04-0.01}$~pb ($<$ 0.22~pb at a 90\% confidence level). 
Using the estimated cross section and a constant decay amplitude approximation, the cross section $\sigma(p\overline{p} \rightarrow
\psi(3770) \pi^0)$ is calculated for the kinematic situation of the planned $\overline{\text{P}}\text{ANDA}$  experiment. 
The maximum cross section corresponding to the two solutions is expected to be less than $0.79$~nb at 90\% confidence level and $122\pm10$~nb at a center of mass energy of 5.26~GeV.
\end{abstract}

\maketitle

\section{Introduction}

The $\overline{\text{P}}\text{ANDA}$  (Anti$\underline{\text{P}}$roton $\underline{\text{An}}$nihilations at $\underline{\text{Da}}$rmstadt) experiment to be built as a part of the 
future \ac{FAIR} located at \ac{GSI} facility in Darmstadt (Germany) will address aspects in the field of non-perturbative quantum chromodynamics \cite{panda_physics_book}. 
The $\overline{\text{P}}\text{ANDA}$ experiment is designed to exploit the physics potential arising from a 
cooled high-intensity antiproton beam covering the center of mass energy range between $\sim$2.3 and 5.5~GeV and
will perform studies of antiproton-proton annihilation and reactions of antiprotons with heavier nuclear targets \cite{panda_physics_book}. The scientific program 
includes among other things the hadron spectroscopy up to the region of charm quarks and especially a detailed investigation
of the spectrum of charmonium and charmonium hybrid states in the open charm sector, including the determination of masses, decay widths, decay properties and quantum numbers~\cite{panda_physics_book}.

All neutral states with non-exotic quantum numbers $J^{PC}$ can be directly produced in $p\overline{p}$ formation reactions. However, $J^{PC}$ exotic states can be produced in association with a meson, 
e.g. an additional pion: $p\overline{p}\rightarrow \pi^0 X$, where $X$ is a $J^{PC}$ exotic hybrid or
charmonium state \cite{PhysRevD.73.096003}. To prepare experiments with $\overline{\text{P}}\text{ANDA}$, estimates for the production cross sections are
required. They can be estimated by models relying on 
constant amplitude approximations and crossing symmetries requiring the a~priori unknown $p\overline{p}\pi^0$ partial decay width of the states as an input parameter \cite{PhysRevD.73.096003}.

$p\overline{p}\pi^0$ decay widths for charmonium states below the open charm threshold have been reported by various
experiments and are relatively well known \cite{pdg_2012}. However, information on the partial decay widths of higher lying charmonium states is still lacking~\cite{pdg_2012}.

The lightest charmonium state which can decay to $D\overline{D}$ pairs is the $\psi(3770)$ resonance. It was predicted by \textit{Eichten et al.}~\cite{PhysRevLett.34.369} 
and discovered by \textit{Rapidis et al.}~\cite{psipp_discovery} and has a mass of $(3773.15\pm 0.33 )$~MeV/c$^2$ and a width of $(27.2\pm1.0)$~MeV \cite{pdg_2012}.

As the mass of the $\psi(3770)$ is slightly above the $D\overline{D}$ threshold and its width is large, it was expected to decay entirely into
$D\overline{D}$ final states \cite{psipp_decay_prediction}. However, the BES collaboration measured the total non-$D\overline{D}$ branching fraction
to be (14.7$\pm$3.2)\% neglecting interference effects \cite{bes_psipp_1,bes_psipp_2,bes_psipp_3,bes_psipp_4}.
The CLEO collaboration measured the non-$D\overline{D}$ branching fraction to be $(-3.3\pm1.4^{+6.6}_{-4.8})$\% taking into account interference 
between electromagnetic resonant and electromagnetic non-resonant (continuum) amplitude assuming no interference with the three-gluon amplitude \cite{cleo_psipp}.
The different results might be explained by different treatments of the interference between electromagnetic resonant and electromagnetic non-resonant (continuum) amplitude.
Meanwhile it has also been noticed that the interference of the continuum amplitude 
with the 3-gluon resonant amplitude, which is dominant compared to the electromagnetic resonant amplitude in $\psi(3770)$ decays~\cite{wang_2}, should be taken into account as well~\cite{wang_1,wang_2}.

The decay channel of $\psi(3770) \rightarrow p\overline{p}$ has been studied recently by the \acs{BESIII} collaboration considering interference between resonant and
continuum amplitude~\cite{bes3_psi3770ppbar}. The measured energy dependence of the cross section was found to be in agreement with destructive interference and two indistinguishable 
solutions for the cross section of $p\overline{p}\rightarrow \psi(3770)$ have been found, one of which is less than 27.5~nb at 90\% confidence level and the other is $425.6^{+42.9}_{-43.7}$~nb.

In this paper, the Born cross section of $e^+e^- \rightarrow p \overline{p} \pi^0$ in the vicinity of the $\psi(3770)$ resonance is studied using data taken by the \ac{BESIII}
experiment. The cross section of the decay $\psi(3770)\rightarrow p \overline{p} \pi^0$  is measured taking into account the interference between the continuum and resonant production amplitudes and
the cross section of $p \overline{p} \rightarrow \psi(3770) \pi^0$, which is an estimate for open charm production in $p\overline{p}$ annihilations,
 for the kinematic situation at the $\overline{\text{P}}\text{ANDA}$ experiment is evaluated using a model based on a constant amplitude approximation \cite{PhysRevD.73.096003}.

\section{Experiment and Data Samples}
The \ac{BESIII} experiment is situated at the \ac{BEPCII} at the Institute of High Energy Physics (IHEP).
 \ac{BEPCII} and \ac{BESIII} \cite{bes_detector} are major upgrades of the BESII experiment and the BEPC collider \cite{bes2_detector}.
They cover the energy range from about $\sqrt{\text{s}}=$2~GeV up to 4.6~GeV and thus allow for the study of physics in the $\tau$-charm energy region.
The double-ring $e^+e^-$ collider is designed for a peak luminosity of 10$^{33}$~cm$^{-2}$s$^{-1}$ at a beam 
current of 0.93~A at the $\psi(3770)$ resonance peak. The detector, with an angular acceptance of about 93\%
of 4$\pi$, consists of 5 major components: (1) The innermost component is a helium gas based \ac{MDC}, which has
in total 43 layers, providing a single wire spatial resolution of 135 micron, a \textit{dE/dx} resolution
of better than 6\%, and a momentum resolution of $\sim$0.5\% for charged particles with momenta of 1~GeV/c 
in a 1~Tesla magnetic field. (2) The \ac{TOF} system for \ac{PID} is composed of a two-layer structure in the barrel and a one layer structure 
in the endcap region. It is built of plastic scintillators and provides a time resolution of 80~ps (110~ps)
in the barrel (endcap) system, allowing a pion/kaon separation at a 95\%~\ac{C.L.} up to about 1~GeV/c. (3)  The \ac{EMC}, which surrounds the \ac{MDC} and the \ac{TOF} system, consists of  
 6240 thallium doped cesium-iodide crystals and provides an energy resolution of 2.5\% (5.0\%) and a position resolution of 6~mm (9~mm) 
for photons with an energy of 1~GeV in the barrel (endcaps) part. 
(4) The superconducting solenoid magnet surrounds these three inner components, providing an axial
uniform magnetic field of 1.0~Tesla. (5) The muon chamber system, embedded in the 
flux return of the magnet, consists of 9 (8) layers of \acp{RPC} in the barrel (endcap)
region and provides a spatial resolution of 2~cm. 

This paper presents a study of $e^+e^-\rightarrow p\overline{p}\pi^0$, which uses the following 
data sets collected with the \ac{BESIII} detector: A data set ($\mathcal{L}=2.9$~fb$^{-1}$) collected at the
peak position of the $\psi(3770)$ resonance (3.773~GeV/c$^2$) \cite{luminosities_arxiv}, a data set ($\mathcal{L}=$44~pb$^{-1}$) collected at a 
center of mass energy of 3.650~GeV \cite{luminosities_arxiv}, and 60~pb$^{-1}$ of data accumulated during a $\psi(3770)$ line-shape
scan. For the sake of statistics, 25 small data sets in the scan data with varying luminosities in the energy range from 3.736~GeV to 3.813~GeV
have been merged together into 7 separate data sets. The resulting center of mass energies of each set have
been calculated by weighting the center of mass energies of the small data sets with their luminosity.
Errors arising from this merging are considered in the systematic error (see section "Systematics
Errors").

A Geant4-based \cite{Agostinelli200325} \ac{MC} simulation software taking into account the geometric and material description of
the \ac{BESIII} detector and the detector response is used for the determination of the detection 
efficiencies, the optimization of event selection criteria and the estimation of backgrounds. 
Since the intermediate products of the decay $\psi(3770)\rightarrow p \overline{p} \pi^0$ (e.g. nucleon resonances) are unknown, 
the detection efficiency has been determined taking into account the kinematic properties
of the decay products at different position in the Dalitz plot using simulated \ac{MC} events.
Polarization effects of the intermediate states, which might affect the detection efficiencies,
are taken into account, too. Therefore, the extracted distribution of the polar angle of the $\pi^0$
in the data, which deviates from a phase space distribution, has been fitted and is taken as input in the simulation of \ac{MC} samples (compare to figure \ref{fig:plot_data_3773}). 
 
\ac{ISR} effects are not considered in the determination of the detection 
efficiencies, but are taken into account later. For the estimation of background contributions from $\gamma_{\text{ISR}}J/\psi$,
$\gamma_{\text{ISR}}\psi(3686)$ and $e^+e^-\rightarrow\psi(3770)\rightarrow~D\overline{D}$  
samples of \ac{MC} events with a size equivalent to 1.3 and 4.8 times of the collected data are analyzed, respectively. Background arising
from exclusive processes similar to the analyzed one, for example decays into $p\overline{p}\pi^0 \gamma$, 
are investigated with \ac{MC} samples containing 20,000 events of the respective process each.

\section{Event Selection}
For $e^+e^- \rightarrow p\overline{p} \pi^0$ events, the $\pi^0$ candidates are reconstructed by their dominant
decay channel into two photons, resulting in a final state with two oppositely charged particles and two
neutral photons. Hence, events with two charged particles resulting in a net charge of zero 
and at least two photon candidates are selected. The polar angles of charged tracks in the \ac{MDC} have to
satisfy $|\cos\theta|{<}0.93$, and the point of closest approach with respect to the $e^+e^-$
interaction point is required to be within $\pm$10~cm in the beam direction and $\pm$1~cm in the plane perpendicular
to the beam axis. The combined information of the specific energy loss of a particle in the \ac{MDC} ($dE/dx$) and its \ac{TOF}
is used to calculate for each charged particle the confidence level for being a pion, a kaon or a proton/antiproton.
A proton/antiproton candidate has to satisfy $CL_{p}{>}CL_{\pi}$ and $CL_{p}{>}CL_{K}$, where $CL_x$ stands
for the respective confidence level of the particle being a proton/antiproton, pion or kaon. Due to differences
in detection efficiencies between data and \ac{MC} simulations for small transverse momenta, 
the proton/antiproton candidates are required to have transverse momenta larger than 300~MeV/c.

Photon candidates are reconstructed by their energy deposition in the \ac{EMC} and are required to 
deposit at least 25~MeV in the barrel region ($|\cos\theta|{<}0.8$) and 50~MeV in the endcaps 
($0.86{<}\cos\theta {<}0.92$). Showers from charged particles in the \ac{EMC}
are suppressed by requiring the angle between a photon candidate and a proton to be larger
than 10$^{\circ}$ and between a photon candidate and an antiproton to be larger than 30$^{\circ}$.

\begin{figure}
\includegraphics[width=0.48\textwidth]{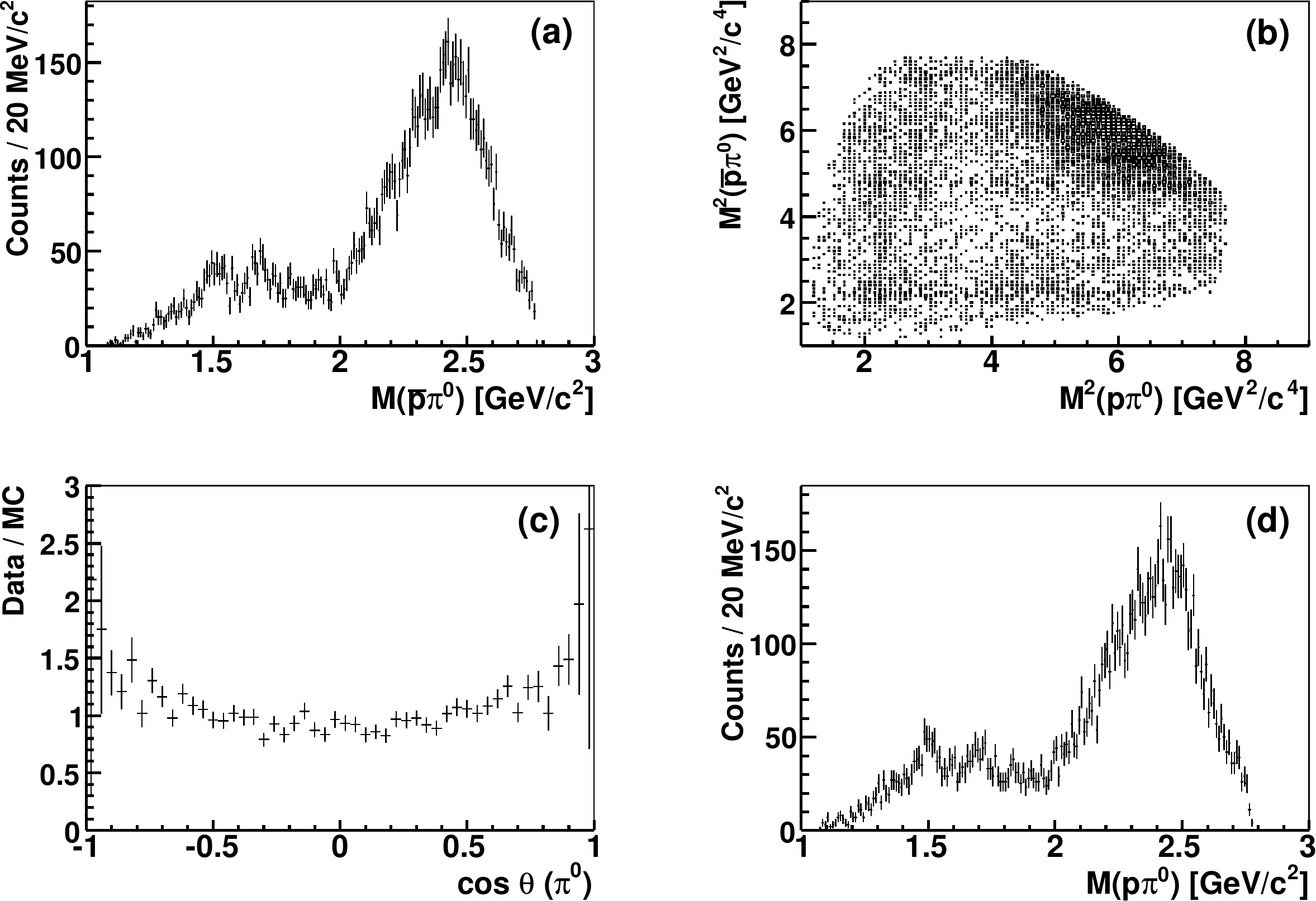}
\caption{\label{fig:plot_data_3773} (a): Distribution of the invariant mass M($\overline{p}\pi^0$);
(b): Dalitz plot of M$^2$($p\pi^0$) vs. M$^2$($\overline{p}\pi^0$); (c): 
ratio of data and phase-space distributed signal \ac{MC} for $\cos\theta$ of $\pi^0$ candidates; and (d): invariant mass M($p\pi^0$), of all candidate events passing the event selection.}
\end{figure}

Events with a proton, an antiproton, and at least two photons are subjected to a five-constraint (5C) kinematic fit
to the initial 4-momentum of the colliding electrons and positrons and to the $\pi^0$ mass of the two photons,
to provide more accurate momentum information on the final state. 
The $\chi^2$ of the kinematic fit is required to be less than 50 and $\pi^0$ candidates are required to be within a 
$\pm$3$\sigma$ region of the $\pi^0$ mass to further reduce background. When more than two photon candidates are found, 
all possible $p\overline{p} \gamma\gamma$ combinations are considered and the one resulting in the smallest $\chi^2$ is 
selected for further analysis. 
Figure \ref{fig:plot_data_3773} shows the Dalitz plot and the projections of the Dalitz plot for all events selected at an energy of 3.773~GeV. 

\section{Background Estimation}
Background from radiative return to the lower lying $J^{PC}=1^{--}$  $\psi(3686)$ and J/$\psi$ resonances, which is not considered in the \ac{ISR}
correction procedure, has been estimated using inclusive \ac{MC} samples. This \ac{ISR} related background is determined to be smaller than 0.5\% 
and its contribution will be considered in the cross section calculation. Contributions from $\gamma _{\text{ISR}}$J/$\psi$ arise mainly due to reconstruction of fake $\pi^0$s
with the radiated \ac{ISR} photons.
Background from $D\overline{D}$ decays at an energy of 3.773~GeV is also estimated using inclusive \ac{MC} samples. 
It is on the order of 0.015\% of all reconstructed events and thus has been neglected. The data taken at 3.650~GeV can contain contributions from
the $\psi(3686)$ tail. Its cross section is estimated in \cite{cross_psi2s_at365} to be $0.136\pm0.012$~nb and this $\psi(3686)$ tail contribution
is also considered in the cross section calculation.
The contributions of other decay channels ($\psi(3770) \rightarrow \gamma \chi_{ci} \rightarrow~\gamma p\overline{p} (\pi^0)\ \text{where}\ i=0,1,2,\ 
\psi(3770)\rightarrow~\gamma \eta_{c} \rightarrow~\gamma p\overline{p}\ ,\   \psi(3770) \rightarrow \gamma \eta_c(2S) \rightarrow \gamma p\overline{p}\ ,\  \psi(3770)\rightarrow p\overline{p}\ ,\  \psi(3770)\rightarrow p\overline{p}\gamma 
\ ,\  \psi(3770)\rightarrow p\overline{p} \pi^0 \gamma\ ,\  \psi(3770)\rightarrow p\overline{p} \pi^0 \gamma \gamma)$ were estimated to be less than 0.4\% of all reconstructed events
at each analyzed energy point. The contributions of decay channels with yet unmeasured branching ratios for the $\psi(3770)$ resonance have been estimated by 
the corresponding decay channels of $\psi(3686)$ and by $\pi^0$ sideband estimations and their number cannot be simply subtracted and will be considered in the systematic error.

\section{Calculation of the Cross Sections}

\begin{table*}
\centering
\caption{Summary of measurements of the number of reconstructed decays into $p\overline{p}\pi^0$ (before applying any corrections), the luminosities $\mathcal{L}$, the average reconstruction efficiency 
$\epsilon_{\text{average}}$ (reconstructed counts divided by efficiency corrected counts), the observed cross sections $\sigma_{obs.}$, the radiative correction factors $1+\delta$ and the calculated Born cross sections $\sigma_{0}$. 
The small efficiency at 3.780~GeV is due to the observed event distribution in the Dalitz plot.
The first error is the statistical error, the second one the \emph{uncorrelated} systematic uncertainty. Correlated systematic uncertainties are not considered here.}\vspace{2.5mm}
\begin{tabular}{ccccccc}
\hline \hline 
Energy   &  \multirow{2}{*}{Counts}& \multirow{2}{*}{$\mathcal{L}$ [pb$^{-1}$]} &  \multirow{2}{*}{$\epsilon_{\text{average}}\ [\%]$} & \multirow{2}{*}{$\sigma_{obs.}^{e^+ e^- \rightarrow  p\overline{p}\pi^0} $ [pb]} & \multirow{2}{*}{$1+\delta$} & \multirow{2}{*}{$\sigma_{0}^{e^+ e^- \rightarrow  p\overline{p}\pi^0} $ [pb]} \\
\text{[GeV]}&  &   &  & & &  \\ \hline
3.650 & $165\pm12.8$     & $44.49\pm 0.02 \pm 0.44 $       &  $43.9 \pm 0.1 \pm1.5$  & $8.44 \pm 0.69 \pm 0.14 $  & 0.84 &  $10.09 \pm 0.84 \pm 0.16$  \\ 
3.746 & $19^{+4.8}_{-4.2}$    & $4.94 \pm 0.01 \pm 0.05$        & $45.5 \pm 0.1 \pm1.5$  & $8.46^{+2.16}_{-1.87} \pm 0.14 $ & 0.88 &  $ 9.60^{+2.45}_{-2.12}  \pm 0.16 $ \\ 
3.753 & $28\pm5.3$       & $9.31 \pm 0.02 \pm 0.10$         & $47.0 \pm 0.1 \pm1.5$ & $6.40 \pm 1.22  \pm 0.10 $ & 0.88 &   $7.28 \pm 1.38 \pm 0.12$\\ 
3.757 & $35\pm5.9$       & $8.04 \pm 0.01 \pm 0.09$          & $47.2 \pm 0.1 \pm1.5$ & $9.22 \pm 1.57 \pm 0.15 $ & 0.88 &   $ 10.44 \pm 1.77\pm 0.17 $ \\ 
3.765 & $42\pm6.5$       & $11.86\pm 0.02 \pm 0.13$          & $45.9 \pm 0.1 \pm1.5$ & $7.72 \pm 1.20 \pm 0.12 $ & 0.88 &   $ 8.73 \pm 1.35 \pm 0.14 $\\ 
3.773 & ~~$9107\pm95.4$~~  & ~~$2916.94 \pm 0.18 \pm 29.17$~~  & ~~$45.5 \pm 0.1 \pm1.5$~~&~~$6.83 \pm 0.08  \pm 0.11 $~~&~~ 0.89 ~~&~~$ 7.71 \pm 0.09 \pm 0.13$~~ \\ 
3.780 & $13^{+4.3}_{-3.7}$       & $5.70  \pm 0.01 \pm 0.06$        & $32.7\pm 0.1 \pm1.5$ & $6.98^{+2.34}_{-2.03}  \pm 0.11 $  & 0.88 &$7.92^{+2.66}_{-2.31} \pm 0.13$ \\ 
3.791 & $45\pm6.7$       & $12.45 \pm 0.02 \pm 0.07$          & $45.8 \pm 0.1 \pm1.5$ & $7.87 \pm 1.18\pm 0.13 $ & 0.87 &$9.03 \pm 1.35 \pm 0.15 $ \\  
3.804 & $33\pm5.8$       & $10.15 \pm 0.03 \pm 0.06$        & $44.2\pm 0.1 \pm1.5$ & $7.37 \pm 1.29   \pm 0.12 $  &0.87 &$ 8.44\pm 1.48 \pm 0.14 $ \\ \hline \hline
\end{tabular}
\label{tab:results_overview}
\end{table*} 

The observed cross sections at $\sqrt{s}$=3.650~GeV and eight more energy points in the range from
3.746~GeV up to 3.804~GeV, including the data collected at the peak of $\psi(3770)$
have been calculated according to $\sigma_{obs.}=\frac{N_{sig.}}{\epsilon \mathcal{L}}$, where $\mathcal{L}$ is
the integrated luminosity, $\epsilon$ the corresponding detection efficiency (which includes also the branching 
fraction of the $\pi^0$ into two photons from \cite{pdg_2012}) and $N_{sig.}$ the number
of events passing the event selection. The background contribution from radiative returns to the lower lying $J^{PC}=1^{--}$  $\psi(3686)$ and J/$\psi$ 
resonances has been subtracted from $N_{sig.}$.

The observed cross section is related to the Born cross section by $\sigma_0=\sigma_{obs.}/(1+\delta)$, with $\sigma_0$ the Born cross section,
$\sigma_{obs.}$ the observed cross section and $1+\delta$ the radiative correction factor, which includes \ac{ISR} contributions,
vertex corrections, and terms arising from the $e^+e^-$ self-energy and the hadronic and leptonic vacuum polarization. 
The factor $1+\delta$ is calculated with the method described in \cite{isr_bes,isr_slac_pub5160}. The contributions from
vertex corrections, $e^+e^-$ self-energy and the hadronic and leptonic vacuum polarization are independent
from the lineshape of the cross section, but not the \ac{ISR} contribution. As input for the lineshape a fit of equation \ref{eq:cross_section} to the observed cross 
sections $\sigma_{obs.}$ is used and the radiative correction factors are calculated. The cross sections are then refined iteratively. At each iteration, the 
radiative correction factors are calculated and the cross sections are updated accordingly. 
After the first iteration, the radiative correction factors change by less than 10\%, after the second iteration by less than 2\%. After the sixth iteration the radiative correction
factors remain constant. The maximum energy for the \ac{ISR} photons considered in the radiative correction procedure is 9\% of the beam energy. 
Table \ref{tab:results_overview} gives an overview of the reconstructed events, determined cross sections, radiative correction factors and
calculated Born cross sections. Figure \ref{fig:cross_section_corrected}
shows the calculated Born cross sections of $e^+e^- \rightarrow p \overline{p} \pi^0$ for the investigated energy points.

\section{Systematic Errors}
Uncorrelated systematic errors in the cross section measurement do not only arise from the aforementioned decay channels (0.4\%, compare to section ``Background Estimation''), but also from the size 
of the \ac{MC} samples (0.5\%) and the efficiency determination. The error on the efficiency has been determined to be smaller than 1.5\% by comparing different 
parametrizations of the Dalitz plot (dividing it into $11\times11$, $22\times22$ and $44\times44$ bins, respectively) for simulated \ac{MC} events. These three error sources will be directly 
considered in the fit to the cross sections (compare to the next section).
 
The dominating systematic error sources are correlated among the different energy points and thus can not be considered directly in the fit --- their effect 
on the final results is estimated by the \textit{offset method} \cite{correlated_errors_1}.
The largest correlated systematic error arises from the radiative correction procedure and the extraction of the Born cross section.
It has been determined by comparing the applied radiative correction procedure with the structure function method proposed 
by \cite{structure_function_1} (2\%), by a comparison of different \ac{MC} decay models (3\%) and by taking the largest difference of the results 
for different cuts (1.5\%).
To take into account polarization effects of intermediate states in the decay, the polar angle of the $\pi^0$ has been extracted by a fit from data. 
The systematic error for this procedure is determined by shifting the extracted values from the fit according to their error and calculating the cross sections again.
The largest difference observed for the cross section is 0.7\% and is taken as the systematic uncertainty.

The error which arises from the finite acceptance of the Dalitz plot at values larger than 7.7~GeV$^2$/c$^4$ on both axes is estimated to be smaller than 2\%.\\
The error from the kinematic fit is due to inconsistencies of the \ac{MC} simulation and data. It is estimated to be less than 2\%
by comparing an inclusive \ac{MC} event sample and a selected control sample of data.
Further errors arise from the \ac{MDC} tracking efficiency (1\% ($p$), 1\% ($\overline{p}$)), the particle identification (1\% ($p$), 2\% ($\overline{p}$)), 
the photon detection efficiency (2\%) \cite{search_for_baryonic_decays} and the error on the center of mass energy measurement of less than 1~MeV.
The integrated luminosity for the data taken at 3.65~GeV and 3.773~GeV was measured using large-angle Bhabha events, and has an 
estimated total uncertainty of 1.1\% \cite{luminosities_arxiv}. The luminosity of the line shape scan data is determined with large angle Bhabha events, too.
The estimated uncertainty is also 1.1\%.

The total systematic uncertainty of the individual energy points is calculated by adding the errors in quadrature and thus is 6.2\%.

To estimate the error of the fit to the Born cross sections (see section ``Fit to the Born Cross Sections''), which arises from the binning of the scan data points, two different sets of binning have
been compared to the nominal one. The largest differences of the central values of the fit are taken a systematic error and are added in quadrature to the systematic error of the fit results. The largest differences
are 16\% (\textit{solution 1}) and 0.7\% (\textit{solution 2}) for the Born cross section $\sigma_{0}(\psi(3770) \rightarrow p\overline{p} \pi^0)$, and 3.7\% (\textit{solution 1}) and 0.1\% (\textit{solution 2})
for the phase~$\phi$.

\section{Fit to the Born Cross Sections}

The cross section of the $\psi(3770)$ decay to $p\overline{p} \pi^{0}$ and other relevant quantities  
are extracted by a fit of 
\begin{eqnarray}
\sigma(s)= & \left| \sqrt{\sigma_{con}} + \sqrt{\sigma_{\psi}} \frac{m\Gamma}{s-m^2+im\Gamma} \exp(i\phi) \right|^2 
\label{eq:cross_section}
\end{eqnarray} 
to the calculated cross sections. The resonant production amplitude is
usually composed of the electromagnetic amplitude and the three-gluon amplitude. However, the electromagnetic amplitude can be neglected 
for the $\psi(3770)$ \cite{wang_2} and thus the resonant production amplitude can be described by $\sqrt{\sigma_{\psi}}$.
The mass $m$ and  the width $\Gamma$ of the $\psi(3770)$ have been fixed according to the world average values \cite{pdg_2012}.
The continuum amplitude $\sqrt{\sigma_{con}}$ can be described by a function of $s$; $\sigma_{con} = C/s^{\lambda}$,
where the exponent $\lambda$ is a priori unknown. The parameter $\phi$ describes the phase for an interference of resonant and
continuum production amplitudes. 

The continuum cross section itself is composed of two different components --- an isospin $I=0$ and an isospin $I=1$
component --- as the  virtual photon arising as an intermediate state in an electron-positron
annihilation can be associated with an isospin of $I=0$ or $I=1$. 
The ratio of the different isospin components of the virtual photon is for example
discussed in \cite{ratio_1_9__1,ratio_1_9__2}. The basic idea is the dominance of single states in the virtual intermediate state, which can be for
example (excited) $\rho^{*}$ or $\omega^{*}$ mesons or coherent pion configurations with $I=0$ and $I=1$, resulting in a
ratio of $({I{=}0}):(I{=}1) = 1:9$. Clearest evidence for such a ratio comes from the process $e^+e^- \rightarrow n\pi$
around 2~GeV, where the ratio of $I=0$ and $I=1$ is measured to be $\sim$1:9 \cite{ratio_1_9__1}. 
Also the electromagnetic decay width $\Gamma_{ee}$ of the $\rho(770)$ and $\omega(782)$ mesons
are in agreement with this ratio \cite{pdg_2012}. A constant ratio at higher energies is also consistent with the ideas of generalized vector
meson dominance \cite{ratio_1_9__1}.

Hence, the continuum amplitude should be expressed as $A_{con}=A_{con}^{I=0}\text{exp}(i\phi_1) + A_{con}^{I=1}\text{exp}(i\phi_2)$ and
Eq.~\ref{eq:cross_section} should contain two different phases. However, they can be combined again as a relative phase and one ends up with
Eq.~\ref{eq:cross_section} considering the interference of the total continuum and resonant amplitudes.

\begin{figure}
\includegraphics[width=0.44\textwidth]{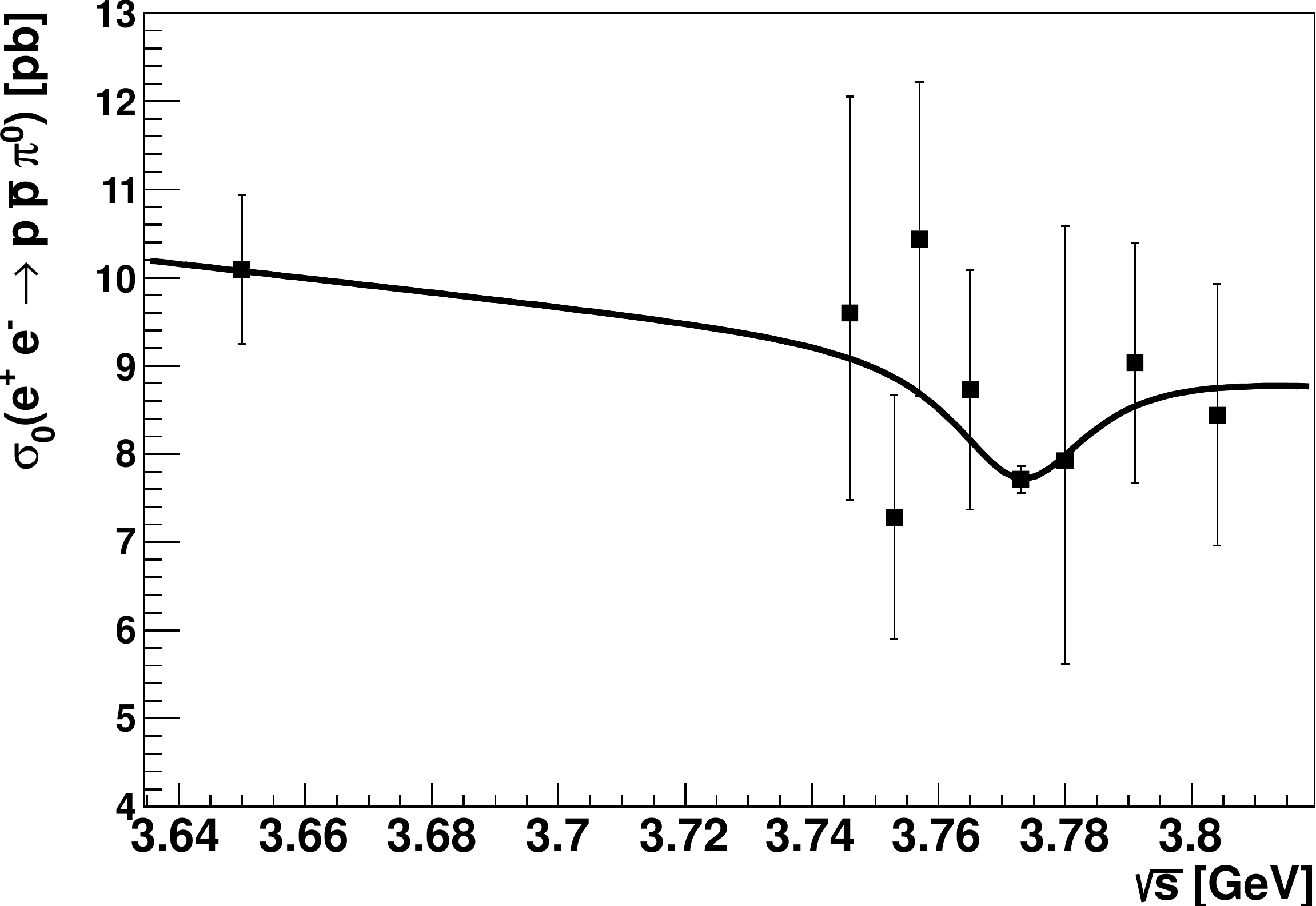}
\caption{\label{fig:cross_section_corrected} The calculated Born cross sections in $e^+e^-$ annihilation to $p\overline{p}\pi^0$ at different 
energy points and the solution of the fitting procedure. The curve of both fit solutions is identical.}
\end{figure}

The maximum likelihood fit yields two different solutions with the same $\chi^2/ndf = 2.70/5 $ and thus with the same probability.
A detailed explanation and a mathematical review of the multiple solution problem is given in \cite{multiple_fit_solution_problem}.
The solutions for the Born cross section $\sigma_{0}(\psi(3770) \rightarrow p\overline{p} \pi^0)$
are $0.06^{+0.10}_{-0.04}$~pb (\textit{solution 1}) and $33.8\pm1.8$~pb (\textit{solution 2}), respectively. 

The results for the phases between resonant and continuum production amplitudes 
are $\phi=269.8^{+52.4}_{-48.0}$~$^{\circ}$ (\textit{solution 1}) and $\phi=269.7\pm 2.3 \ ^{\circ}$ (\textit{solution 2}), which both are in agreement with a destructive interference ($\hat{=}270^{\circ}$).
The parameters describing the slope of the continuum cross section are $C=(0.4\pm0.1)\cdot10^{3}\text{ GeV}^{2\lambda}\text{pb}$ (\textit{solution 1}), $C=(0.4\pm0.6)\cdot10^{3}\text{ GeV}^{2\lambda}\text{pb}$
(\textit{solution 2}) and $\lambda=1.4\pm0.1$ (\textit{solution 1}) and $\lambda=1.4\pm0.6$ (\textit{solution 2}). The unit and the large error of $C$ are arising from the correlation of $C$ and $\lambda$. 
The fit is shown in Figure \ref{fig:cross_section_corrected}. 
The statistical significance of the resonant amplitude, calculated based on the differences of the likelihood values between the fit with and without
assuming a resonant contribution, is for both solutions 1.5$\sigma$. 

Using the estimated cross section and the Born cross section of the $\psi(3770)$ resonance as given in \cite{search_for_baryonic_decays}, the cross section
of the processes $p\overline{p}\rightarrow \psi(3770) \pi^0$ can be calculated based on a constant decay amplitude approximation 
\cite{PhysRevD.73.096003} for the kinematic situation at the $\overline{\text{P}}\text{ANDA}$ experiment. According to this model the maximum cross section
can be expected at a center of mass energy of 5.26~GeV, which is still in reach of $\overline{\text{P}}\text{ANDA}$ \cite{panda_physics_book}; 
the results are $<0.79$~nb at a 90\% confidence level (\textit{solution 1}) and $122\pm10$~nb (\textit{solution 2}). The error for \textit{solution 2} has been determined
by taking the difference of the original solution and the solution using the sum of measured value and error as input value for the calculation.

Table \ref{tab:final_results_overview} shows a compilation of the obtained results. Here, the systematic errors are considered. 
The upper limits at 90\% \ac{C.L.} are given for parameters where the error is dominating the measurement. The upper limits have been calculated assuming
that the statistical and systematic errors are following a bifurcated Gaussian distribution.

\section{Summary}
Using 2.9~fb$^{-1}$ of data collected at $\sqrt{s}=3.773$~GeV, 44~pb$^{-1}$ of data collected at $\sqrt{s}=3.650$~GeV
and 60~pb$^{-1}$ of data collected during a $\psi(3770)$ line shape scan with the \ac{BESIII} detector at the
\ac{BEPCII} collider, an analysis of the process $e^+e^-\rightarrow p\overline{p}\pi^0$ has been performed. The Born cross section
of $e^+e^-\rightarrow \psi(3770) \rightarrow p\overline{p}\pi^0$ has been extracted allowing the continuum production amplitude to
interfere with the resonant production amplitude. Two solutions with the same probability are found. 
The Born cross section is determined to be $0.06^{+0.10+0.01}_{-0.04-0.01}$~pb ($<$~0.22~pb at a 90\% C.L.) or $33.8\pm 1.8 \pm 2.1 $~pb. 
 Both phases of the fit solution are consistent with $270^{\circ}$, which is in agreement with a destructive interference.
Using a constant decay amplitude approximation, the cross sections of $p\overline{p}\rightarrow \psi(3770) \pi^0$ are calculated to be
less than $0.79$~nb at a 90\% \ac{C.L.} and $122\pm10$~nb at center of mass energy of 5.26~GeV \cite{PhysRevD.73.096003}, respectively.

\begin{table}
 \centering
 \caption{A summary of the extracted results from the fit. The upper limits are determined at a 90\%
 \ac{C.L.}, where the first error given is from the fit (i.e. from the uncorrelated sources) and the second error is from the correlated systematics.
 The phases given are the one from the solutions of the fitting procedure.}\vspace{2.5mm}
 \renewcommand{\arraystretch}{1.2}
 \begin{tabular}{c|c|c|c}
 \hline \hline 
  \multirow{2}{*}{Solution}   & $\sigma^{\psi(3770) \rightarrow p\overline{p} \pi^0}_{0}$ & \multirow{2}{*}{$\phi_{Fit}$ [$^{\circ}$]} & $\sigma_{0}^{p\overline{p} \rightarrow \psi(3770) \pi^0}$  \\
            & [pb] &   &  [nb] at 5.26~GeV  \\ \hline 
   (1)  & $<$ 0.22  & $269.8^{+52.4}_{-48.0}{\pm}11.0$ & $<0.79$ \\ 
   (2)  & $33.8{\pm}1.8{\pm}2.1$ & $269.7{\pm}2.3{\pm}0.3$  & 122${\pm}$10 \\ 
  \hline  \hline
 
\end{tabular}
\label{tab:final_results_overview}
\end{table}

\section{Acknowledgement}
The BESIII collaboration thanks the staff of BEPCII and the computing center for their strong support. This work is supported in part by the 
Ministry of Science and Technology of China under Contract No. 2009CB825200; Joint Funds of the National Natural Science Foundation of China 
under Contracts Nos. 11079008, 11179007, U1332201; National Natural Science Foundation of China (NSFC) under Contracts Nos. 10625524, 10821063, 10825524, 
10835001, 10935007, 11125525, 11235011; the Chinese Academy of Sciences (CAS) Large-Scale Scientific Facility Program; CAS under Contracts Nos. KJCX2-YW-N29, KJCX2-YW-N45; 
100 Talents Program of CAS; German Research Foundation DFG under Contract No. Collaborative Research Center CRC-1044; Istituto Nazionale di Fisica Nucleare, Italy; 
Ministry of Development of Turkey under Contract No. DPT2006K-120470; U. S. Department of Energy under Contracts Nos. DE-FG02-04ER41291, DE-FG02-05ER41374, 
DE-FG02-94ER40823, DESC0010118; U.S. National Science Foundation; University of Groningen (RuG) and the Helmholtzzentrum fuer Schwerionenforschung GmbH (GSI), Darmstadt; 
WCU Program of National Research Foundation of Korea under Contract No. R32-2008-000-10155-0.
\bibliography{bibliothek}

 \begin{acronym}[SQL]
  \acro{MC}{Monte Carlo}
  \acro{C.L.}{confidence level}
  \acro{BESIII}{Beijing Spectrometer III}
  \acro{BEPCII}{Beijing Electron-Positron Collider II}
  \acro{IHEP}{Institute of High Energy Physics}
  \acro{MRPC}{Multigap Resistive Plate Chambers}
  \acro{TOF}{Time of Flight}
  \acro{MDC}{Main Drift Chamber}
  \acro{ISR}{initial state radiation}
  \acro{EMC}{Electromagnetic Calorimeter}
  \acro{SSM}{Superconducting Solenoid Magnet}
  \acro{PID}{particle identification}
  \acro{BOSS}{BESIII Offline Software System}
  \acro{RPC}{Resistive Plate Chamber}
  \acro{ISR}{Initial State Radiation}
\acro{FAIR}{Facility for Antiproton and Ion Research}
\acro{GSI}{Gesellschaft f\"ur Schwerionenforschung}

 \end{acronym}

\end{document}